\documentclass[aps,pre,twocolumn,groupedaddress,showpacs]{revtex4-1}
\newcommand{\bec}[1]{\mbox{\boldmath $ #1$}}
\usepackage{graphicx}
\usepackage{bm}
\begin{document}
\title{Acceleration of raindrops formation due to tangling-clustering
instability in turbulent stratified atmosphere}
\author{T. Elperin$^1$}
\email{elperin@bgu.ac.il}
\homepage{http://www.bgu.ac.il/me/staff/tov}
\author{N. Kleeorin$^{1}$}
\email{nat@bgu.ac.il}
\author{B. Krasovitov$^1$}
\email{borisk@bgu.ac.il}
\author{M. Kulmala$^2$}
\email{markku.kulmala@helsinki.fi}
\author{M. Liberman$^{3,4}$}
\email{misha.liberman@gmail.com}
\homepage{http://michael-liberman.com/}
\author{I.~Rogachevskii$^{1}$}
\email{gary@bgu.ac.il}
\homepage{http://www.bgu.ac.il/~gary}
\author{S. Zilitinkevich$^{2,5,6,7}$}
\email{sergej.zilitinkevich@fmi.fi}

\medskip
\affiliation{$^1$The Pearlstone Center for
Aeronautical Engineering Studies, Department of Mechanical Engineering,
Ben-Gurion University of the Negev, P. O. Box 653, Beer-Sheva
84105, Israel \\
 $^2$Division of Atmospheric Sciences, Department of Physics,
P.O. Box 64, 00014 University of Helsinki,
Finland \\
 $^3$Nordita, KTH Royal Institute of Technology
and Stockholm University, Roslagstullsbacken 23,
10691 Stockholm, Sweden \\
 $^4$Moscow Institute of Physics and Technology,
Dolgoprudnyi, 141700, Russia \\
 $^5$Finnish Meteorological Institute (FMI) PO Box 503,
00101 Helsinki, Finland \\
 $^6$Department of Radio Physics, N.~I.~Lobachevsky State University of
Nizhny Novgorod, Russia \\
 $^7$Moscow State University; Institute of Geography of Russian Academy of Sciences,
Moscow, Russia
}

\date{\today}
\begin{abstract}
Condensation of water vapor on active cloud
condensation nuclei produces micron-size water
droplets. To form rain, they must grow rapidly
into at least 50-100 $\mu$m droplets.
Observations show that this process takes only
15-20 minutes. The unexplained physical mechanism
of such fast growth, is crucial for understanding
and modeling of rain, and known as
"condensation-coalescence bottleneck in rain
formation". We show that the recently discovered
phenomenon of the tangling clustering instability
of small droplets in temperature-stratified
turbulence (Phys. Fluids 25, 085104, 2013)
results in the formation of droplet clusters with
drastically increased droplet number densities.
The mechanism of the tangling clustering
instability is much more effective than the
previously considered by us the inertial clustering instability
caused by the centrifugal effect of turbulent
vortices. This is the reason of
strong enhancement of the collision-coalescence rate inside the
clusters. The mean-field theory of the droplet growth developed in this study can
be useful for explanation of the observed fast growth of cloud
droplets in warm clouds from the initial 1 $\mu$m size droplets
to 40-50 $\mu$m size droplets within 15-20
minutes.
\end{abstract}


\maketitle

\section{Introduction}

When ascending parcel of moist air reaches the
condensation level, the initial mist of small,
micron-size water droplets is formed, which are
suspended in the air. In the super-saturated
environment water droplets grow due to
condensation of water vapor from the surrounding
atmosphere. However, to form the raindrops, which
can fall down triggering rain, they must grow up
to about 50 $\mu$m size droplets, which would
take a very long time. Observations indicate that
the average time for rainfall initiation is
approximately $15 - 20$ minutes, while existing
theories predict that the duration of a time
interval, required for droplets to grow up to 50
$\mu$m in radius, is of the order of hours (see,
e.g., reviews~\cite{Khain-et-al2007,Devenish-et-al2012,Grabowski-and-Wang2013},
and references therein). Indeed, though the
actual time of large droplets formation depends
on the initial droplet size spectrum and cloud
water content (see, e.g.,
\cite{Pruppacher-and-Klett1997}), the predicted
growth time differs considerably from the
observations.

Initiation of rain in turbulent clouds comprises
three stages. The first stage involves
condensation of water vapor on cloud condensation
nuclei (CCN, typically having a size of the order
of 0.05 $\mu$m) and formation of small micron
size droplets. At the next stage, droplets grow
efficiently through condensation and diffusion of
water vapor and may attain radii of about 10
$\mu$m. It is generally believed that droplets
having radii larger than 50 $\mu$m fall out of
the cloud due to gravitational sedimentation and
continue to grow in size mainly through
gravitational collisions into rain droplets with
the size of the order of $80 - 100$ $\mu$m.
Understanding a mechanism of rapid growth of
initially small droplets to the size of the order
of 50 $\mu$m when gravitational
collision-coalescence becomes effective is still
poorly understood and remains a subject of active
research
(see, e.g., \cite{Khain-et-al2007,Devenish-et-al2012,Grabowski-and-Wang2013}).
Identifying mechanisms of rapid growth of cloud
droplets and determining the growth rate, i.e.
theoretical explanation of the so-called
"\textit{size gap or the condensation-coalescence
bottleneck in warm rain formation}"
\cite{Grabowski-and-Wang2013} is one of the
major challenges in cloud physics.

Observations show the existence of strong
turbulence in clouds. Different mechanisms have
been suggested and different aspects of
turbulence effects on the growth of cloud
droplets have been considered to explain the
rapid formation of rain droplets in clouds
\cite{Grabowski-and-Wang2013}. These mechanisms
involve e.g. effects of giant aerosol particles
for faster formation of large cloud droplets,
thereby initiating coalescence sooner
\cite{Blyth-et-al2003} and droplet spectra
broadening under conditions of water vapor
supersaturation
\cite{Vaillancourt-et-al2001,Vaillancourt-et-al2002,Tisler-et-al2005}.
Numerous theoretical, numerical and experimental
studies used different approaches and models to
investigate the effects of atmospheric turbulence
on growth of cloud droplets by
collision-coalescence and formation of rain
droplets
(see \cite{Khain-et-al2007,Devenish-et-al2012,Grabowski-and-Wang2013},
and references therein).

Most of the studies have focused on amplification
of the fall velocity of cloud droplets in
turbulent atmosphere and turbulence induced
increase of the droplet collision kernel. Air
turbulence can enhance droplet coalescence rate
by increasing the relative velocity of droplets
due to differential acceleration and enhance
collision kernel of cloud droplets. For example,
when the dissipation rate of turbulence is increased from
100 to 400 cm$^2$ s$^{-3}$, the droplet
coalescence rate (between droplets with the sizes
18 $\mu$m and 20 $\mu$m) increases by a factor of
3.5 \cite{Wang-et-al2008}. The increase of droplet relative velocity
and local accumulation of inertial droplets near
the periphery of turbulent eddies due to
centrifugal effect, can increase droplet
collision rate (see, e.g.,  \cite{Khain-et-al2007,Wang-et-al2008,Wang-and-Maxey1993,Elperin-et-al1996b,Elperin-et-al2002,Elperin-et-al2007,Pinsky-and-Khain1997a,Pinsky-and-Khain1997b,Pinsky-and-Khain2002,Pinsky-and-Khain2004,Wang-et-al2000,Wang-et-al2005,Zhou-et-al2001,Davila-and-Hunt2001,Dodin-and-Elperin2002,Falkovich-et-al2002,Ghosh-et-al2005,GM2015}).
Numerical simulations showed that due to the
effect of preferential concentration of inertial
particles in turbulent flows their settling rate
is about 20\% \ larger than the terminal fall
velocity in the quiescent atmosphere (see, e.g.,
 \cite{Khain-et-al2007,Wang-and-Maxey1993,Pinsky-and-Khain1997a,Pinsky-and-Khain1997b,Wang-et-al2000,Zhou-et-al2001,Davila-and-Hunt2001,Squires-and-Eaton1991}).
This list of references is obviously not complete
because the topic is a subject of intense ongoing
research and attracts attention of numerous
researchers (see, e.g.,
\cite{Devenish-et-al2012,Grabowski-and-Wang2013}).

Accurate modeling of droplet
collision-coalescence is important because
collisions strongly affect droplet size and
velocity distributions, and dispersion of
droplets \cite{Gavaises-et-al1996}. Droplet
collisions may have numerous outcomes - the
droplets might smoothly merge with little
deformation, bounce off each other, coalesce
following large deformation, or separate after
temporarily coalescing. Many of the used droplet
interaction models assume that droplet velocities
before collisions are not correlated. However,
this assumption is violated in turbulent flows.
Indeed, small droplets have low inertia and
follow almost the same trajectories as fluid
particles and, therefore, their pre-collision
velocities are strongly correlated with the
velocity of a carrying fluid
\cite{Villedieu-and-Simonin2004}. Many of the
studies focused on collisions between identical
droplets whereby the collision outcome depends on
the impact parameter and the ratio of kinetic
energy to surface tension. It was demonstrated
that size disparity can significantly increase
the parameter range over which droplets
permanently coalesce \cite{Tang-et-al2012}.

Dynamics and interactions of liquid droplets,
their collisions, coalescence and bouncing,
become more significant with increase of their
size and are encountered in many naturally
occurring phenomena and industrial applications,
including rain initiation and combustion.
Nevertheless, the collision rate for typical
droplet number densities in clouds is too far
from being sufficient for their efficient
coalescence. The general opinion is that
turbulence somehow enhances droplet collision
rate and droplet coalescence. However, it still
remains unclear and not completely established
yet to what extent and how turbulence can affect
and control droplet coalescence and rain
initiation
(see, e.g., \cite{Devenish-et-al2012,Grabowski-and-Wang2013}).

In this paper we explain the fast growth of cloud
droplets by collision-coalescence taking into
account recently discovered phenomenon of
tangling clustering instability of small water
droplets in turbulent temperature stratified
atmosphere \cite{Elperin-et-al2013}. We assume
that water droplets coalesce after collisions.
However, the ambient mean number density of the
droplets is too low, so that their
collision-coalescence time is very large. The
situation dramatically changes in the presence of
tangling clustering instability which results in
the formation of clusters with the mean number
density of the droplets inside the clusters that
by several orders of magnitude exceeds the
ambient mean number density of the droplets.

The mechanism of droplet clustering in turbulence
is as follows. Due to inertia effects droplets
inside turbulent eddies are carried out to the
boundary between the eddies by inertial forces.
Therefore, water droplets are locally accumulated
in the regions with low vorticity and maximum
pressure fluctuations \cite{Maxey1987}. Contrary
to the inertia induced preferential
concentration, the pressure fluctuations in
stratified turbulence with a nonzero mean
temperature gradient are increased due to
additional temperature fluctuations generated by
tangling of the mean temperature gradient by
velocity fluctuations. This is a reason why
clustering of water droplets is much more
effective in stratified turbulence
\cite{Elperin-et-al2013,Eidelman-et-al2010} in
comparison with a non-stratified turbulence
\cite{Elperin-et-al2002}.

The tangling clustering instability leads to the
formation of clusters, which accumulate
surrounding droplets. Since the number density
and, correspondingly, the collision-coalescence
rate of small droplets inside the clusters
drastically increase, the characteristic time of
droplet coalescence sharply decreases. Effect of
the tangling clustering instability
\cite{Elperin-et-al2013} is much stronger than
that of the inertial clustering instability
\cite{Elperin-et-al2002} in non-stratified
isotropic and homogeneous turbulence. The
strong enhancement of droplet
collision-coalescence rate caused by the effect
of tangling clustering instability of small
droplets, can
explain the observed fast growth of cloud
droplets from the initial 1 $\mu$m size droplets
to 40-50 $\mu$m size droplets within 15-20
minutes.

\section{Tangling clustering instability }

Small cloud droplets with the size of the order
of 1 $\mu$m have to grow in diameter by a factor
50-100 in order to fall out of the cloud as rain
droplets. Initial formation of cloud droplets is
associated with an intricate process that allows
conversion of water vapor into small liquid water
droplets. Droplet formation always requires the
presence of aerosols and their activation to
cloud droplets, and further growth of droplets
via condensation-coalescence. Clearly, the growth
of cloud droplets is constrained by their
vaporization, and droplet collisions and
coalescence may modify the droplet size
distribution (see, e.g.,
\cite{Eggers1997,Eggers-et-al1999,Duchemin-et-al2003}).

In the present study we invoke recently
discovered phenomenon of tangling clustering
instability of droplets in temperature stratified
turbulence which causes formation of clusters
with the droplet number density inside the
clusters by several orders of magnitude larger
than the ambient droplet number density
\cite{Elperin-et-al2013}. The size of the formed
clusters is of the order of the Kolmogorov
micro-scale length. For the sake of simplicity in
this section we assume that vapor condensation
produces small droplets of the same size, which
then grow due to the collision-induced
coalescence. The droplet size distribution is
taken into account in the next section.

\subsection{Governing equations}

The theory of the tangling clustering
instability in the temperature-stratified
turbulence has been developed in \cite{Elperin-et-al2013}.
In this section we summarize these theoretical results
and explain why the clustering instability is
essentially enhanced in the turbulence
with large-scale temperature gradient.
Equation for the instantaneous number
density $n(t,{\bm r})$  of small spherical droplets
in a turbulent flow reads:
\begin{eqnarray}
\frac{\partial n}{\partial t} + {\bm \nabla
\cdot} \left(n \,{\bm v}\right) = D_m\, \Delta n
- \frac{n}{\tau_{ev}} + I_0,
 \label{Eq1}
\end{eqnarray}
where $D_m = k_B \,T/(3 \pi \rho \, \nu \, d)$ is
the coefficient  of  molecular  (Brownian)
diffusion of droplets having the diameter $d$ and
the instantaneous velocity ${\bm v}(t,{\bm r})$, $\nu$  is the kinematic viscosity, $T$
and $\rho$  are the mean air temperature and
density, $k_B$ is the Boltzman constant and $I_0$  is the rate of production
of the droplets number density caused by an external source of droplets, e.g.,
through activation.
The term $- n/\tau_{ev}$ in the right hand side of Eq.~(\ref{Eq1})
describes the decrease of the droplet number density
due to the evaporation, where $\tau_{ev}$ is the characteristic evaporation
time determined by Eq.~(\ref{Eq8}) in Sect. IIIB, see, e.g., \cite{Nadykto-et-al2003}.

The droplet velocity ${\bm v}$ is
determined by the equation of motion:
\begin{eqnarray}
{d{\bm v} \over dt} = {{\bm u} - {\bm v} \over
\tau_{\rm st}} + {\bm g} .
 \label{Eq3}
\end{eqnarray}
Here ${\bm u}(t, {\bm x})$ is the fluid velocity and
${\bm g}$ is the gravity acceleration,
$\tau_{\rm st}=m_{\rm dr} /3 \pi \rho \, \nu \,d$ is the Stokes
time, $m_{\rm dr} = \left(\pi /6\right)\, \rho_m
\, d^3$ is the droplet mass, and $\rho_m \gg
\rho$ is the droplet mass density. The ratio,
${\rm St} =\tau_{\rm st}/{\tau}_{\eta} = \rho_{m}
\, d^2 /18\rho \, \ell_\eta^2$, of the Stokes
time and the Kolmogorov turbulent turnover time,
$\tau_\eta$, is the Stokes number, where
$\tau_\eta= \ell_\eta /u_\eta = \tau _0/{\rm Re}^{1/2}$,
$u_\eta=u_0/{\rm Re}^{1/4}$ is the
characteristic velocity of eddies in the
Kolmogorov micro-scale, $\ell_\eta = \ell _0/{\rm
Re}^{3/4}$, ${\rm Re} = u_0 \ell _0/\nu$ is the
Reynolds number, $u_0$ is the characteristic
turbulent velocity in the integral turbulent
scale $\ell_0$ and $\tau_0 = \ell_0/u_0$ is the
turbulent time in the integral turbulent scale.

Solution of Eq.~(\ref{Eq3}) for ${\rm St} \ll 1$ reads (see, e.g.,
\cite{Maxey1987}):
\begin{eqnarray}
{\bm v} = {\bm u} - \tau_{\rm st}\, \biggl[{\partial
{\bm u} \over \partial t} + ({\bm u} {\bm \cdot}
{\bm\nabla}) {\bm u} - {\bm g}\biggr] + {\rm
O}(\tau_{\rm st}^2) .
 \label{ED2}
 \end{eqnarray}
This equation implies that ${\bm \nabla} {\bm \cdot} \, {\bm v} \not=0$, i.e.,
the droplet velocity field is compressible,
\begin{eqnarray}
{\bm \nabla} {\bm \cdot} \, {\bm v} &=& {\bm
\nabla} {\bm \cdot} \, {\bm u} - \tau_{\rm st} \, {\bm
\nabla} {\bm \cdot} \,  \biggl( {d{\bm u} \over
dt} \biggr) + {\rm O}(\tau_{\rm st}^2)
\nonumber\\
&=& - {1 \over
\rho} \, ({\bm u} {\bm \cdot} {\bm\nabla})  \rho
+ {\tau_{\rm st} \over \rho} \,{\bm\nabla}^2 p  + {\rm
O}(\tau_{\rm st}^2) .
 \label{ED3}
\end{eqnarray}
In derivation of Eq.~(\ref{ED3}) we used the Navier-Stokes equation for the fluid.
The mechanism of the clustering instability is associated with the droplet inertia.
The centrifugal forces cause the droplets inside the turbulent eddies drift out
to the boundary between the eddies, i.e., to the regions with the maximum
fluid pressure fluctuations. Indeed,  for a large Peclet number,
when the molecular diffusion of  droplets in
Eq.~(\ref{Eq1}) can be neglected, we can estimate
$dn / d t   \propto - {\bm \nabla} {\bm \cdot} \, {\bm v}$.
Here we neglected evaporation and consider the case $I_0=0$.
Since ${\bm \nabla} {\bm
\cdot} \, {\bm v} \propto (\tau_{\rm st} /\rho)
\,{\bm\nabla}^2 p \not=0$ even for
incompressible fluid, this implies that $dn / dt \propto - (\tau_{\rm st}
/\rho) \,{\bm\nabla}^2 p > 0$ in the regions where ${\bm\nabla}^2 p < 0$.
Therefore, the droplets are accumulated in regions with
maximum pressure fluctuations.

Averaging Eq.~(\ref{Eq1}) over an ensemble of turbulent velocity field
we obtain the equation for the mean number density of droplets $N =\langle n \rangle$
\begin{eqnarray}
\frac{\partial N}{\partial t} + {\bm \nabla
\cdot} \left(N \,{\bm V}_p + \langle n' \,{\bm v}'\rangle \right)
= D_m \, \Delta N - \frac{N}{\tau_{ev}} + I_0,
 \label{Eq1a}
\end{eqnarray}
where ${\bm v}'$ and $n'$ are the fluctuations of the droplet
velocity and number density, respectively,
${\bm V}_p$ is the mean droplet velocity that is
the sum of the mean fluid velocity, ${\bm U}$,
and the terminal fall velocity  of droplets, ${\bm V}_g={\bm g}\tau_{\rm st}$
[see Eq.~(\ref{ED2})].

The clustering instability of droplets in turbulent flow is determined  by
fluctuations of the droplet number
density, $n'(t,{\bm r})= n(t,{\bm r}) - N(t,{\bm
r})$. Equation for the fluctuations $n'$ is obtained by
subtracting Eq.~(\ref{Eq1a}) from Eq.~(\ref{Eq1})
\cite{Elperin-et-al2013}:
\begin{eqnarray}
\frac{\partial n'}{\partial t} &+& {\bm \nabla
\cdot} \left[n' \,\left({\bm v}'+{\bm V}_p \right) - \langle n' \,{\bm
v}'\rangle \right] - D_m\, \Delta n' = - ({\bm v}' {\bm
\cdot \nabla}) N
\nonumber\\
&-& N \, {\bm \nabla \cdot} \,{\bm
v}' - \frac{n'}{\tau _{ev}}.
 \label{Eq2}
\end{eqnarray}

\subsection{Mechanism of tangling clustering
instability}

In a case of temperature stratified
turbulence with a non-zero large-scale temperature gradient,
the turbulent heat flux $\langle {\bm
u}' \, \theta \rangle$ is not zero, where ${\bm
u}'$ are the fluctuations of the fluid velocity. This
implies correlation between fluctuations of fluid
temperature, $\theta$, and velocity, and, therefore,
the correlation between fluctuations of
pressure and fluid velocity.
In a temperature stratified
turbulence there are additional pressure fluctuations
caused by the tangling of the mean temperature gradient
by the velocity fluctuations. This causes the
increase of pressure fluctuations, and correspondingly enhance
the droplet clustering.
The tangling clustering mechanism is dynamically similar to the inertial clustering
mechanism. In particular, the inertial particles drift out to the regions with higher
pressure fluctuations, i.e., the regions with lower vorticity and
higher strain rate. However, in the temperature stratified turbulence
the pressure fluctuations are stronger than in non-stratified turbulence.
Since the clustering is related to the Laplacian of the pressure
[see Eq.~(\ref{R1}) below]
this is the reason of the enhanced tangling clustering.

Fluctuations of the droplet number density
are described by
the two-point second-order correlation function,
$\Phi(t,{\bm R})=\langle n'(t,{\bm x}) n'(t,{\bm
x}+{\bm R}) \rangle$.
The analysis of the tangling clustering instability
employs the equation for
the correlation function $\Phi(t,{\bm R})$ that has been derived
using the path-integral approach for random
compressible flow with a finite correlation time
\cite{Elperin-et-al2002}:
\begin{eqnarray}
{\partial \Phi \over \partial t} &=& \biggl[B({\bm
R})-{2 \over \tau _{ev}} + 2 {\bm U}^{(A)}({\bm R})\cdot {\bm\nabla}
\nonumber\\
&& \quad +
\hat D_{ij}({\bm R}) \nabla_{i} \nabla_{j}\biggr]
\, \Phi(t,{\bm R}),
 \label{S2}
\end{eqnarray}
where ${\bm U}^{(A)}({\bm R}) = (1/2) \,
\big[\tilde {\bm U}({\bm R}) - \tilde{\bm U}(-{\bm R})\big]$,
\begin{eqnarray}
B({\bm R}) &\approx& 2 \int_{0}^{\infty} \langle
b\big[0,{\bm\xi}(t,{\bm x}|0)\big]
\,b\big[\tau,{\bm\xi}(t,{\bm x}+{\bm
R}|\tau)\big] \rangle \,d \tau ,
 \label{S6}\\
\tilde U_{i}({\bm R}) &\approx& -2 \int_{0}^{\infty}
\langle v'_{i}\big[0,{\bm\xi}(t,{\bm x}|0) \big]
\,b\big[\tau,{\bm\xi}(t,{\bm x}+{\bm
R}|\tau)\big] \rangle \,d \tau ,
\nonumber\\
 \label{S7}\\
 \hat D_{ij}&=& 2 D_m \delta_{ij} +
D_{ij}^{^{T}}(0) - D_{ij}^{^{T}}({\bm R}) ,
 \label{S4}\\
D_{ij}^{^{T}}({\bm R}) &\approx& 2
\int_{0}^{\infty} \langle v'_{i}
\big[0,{\bm\xi}(t,{\bm x}|0)\big] \,
v'_{j}\big[\tau,{\bm\xi}(t,{\bm x}+{\bm
R}|\tau)\big] \rangle \,d \tau .
\nonumber\\
\label{S5}
\end{eqnarray}
Equation~(\ref{S2}) is written in the frame moving with the mean droplet velocity.
The function $B({\bm R})$ is
determined by the compressibility of the droplet
velocity field, $b={\rm div} \,{\bm v}'$. The vector $\tilde {\bm U}({\bm R})$
determines a scale-dependent drift velocity which
describes transport of fluctuations of droplet
number density from smaller scales to larger
scales. The tensor of the scale-dependent turbulent diffusion
$D_{ij}^{^{T}}({\bm R})$ tends to the tensor of the molecular
(Brownian) diffusion at very small scales, while
in the vicinity of the integral turbulent
scale it coincides with the tensor of turbulent
diffusion.
Other variables in Eqs.~(\ref{S2})-(\ref{S5}) are defined as follows:
$\delta_{ij}$ is the Kronecker
tensor, the Wiener trajectory ${\bm\xi}(t,{\bm
x}|s)$ in the expressions for the turbulent
diffusion tensor $D_{ij}^{^{T}} ({\bm R})$ and
other transport coefficients is:
${\bm\xi}(t,{\bm x}|s) ={\bm x} - \int^{t}_{s}
{\bm v} [\tau,{\bm\xi}(t,{\bm x}|\tau)] \,\,d
\tau + \sqrt{2 D_m} \, {\bm w}(t-s)$, and $\langle ... \rangle$ denotes
averaging over the statistics of turbulent
velocity field and the Wiener random process ${\bm
w}(t)$ that describes the Brownian motion.
The second term in the right hand side of Eq.~(\ref{S2})
describes the effect of droplets evaporation.

The exponential growth of the correlation function
of the droplet number density fluctuations,
$\Phi(t,{\bm R})$, due to the tangling clustering instability,
is determined by the first term, $B({\bm R}) \, \Phi(t,{\bm R})$,
in the right hand side of Eq.~(\ref{S2}), which is the only positive one.
To estimate the function $B({\bm R})$ we
take into account the equation of state of an
ideal gas that yields: $p'/P = \rho'/\rho +
\theta/T + O(\rho' \theta/\rho T)$, where
$\rho, T, P$ and $\rho', \theta, p'$ are
the mean and fluctuations of the fluid density,
temperature, and pressure, respectively. For small
Stokes numbers, ${\bm \nabla} {\bm \cdot} \, {\bm
v}' \approx (\tau_{\rm st}/\rho) \, {\bm\nabla}^2 p' + O({\rm St}^2)$,
we obtain
\begin{eqnarray}
B({\bm R}) &\approx& {2 \tau_{\rm st}^2 \over \rho^2} \,
\langle  \tau \big[{\bm\nabla}^2 p'({\bm x})
\big]\, \big[{\bm\nabla}^2 p'({\bm y})\big]
\rangle
\nonumber\\
& \approx &{2 \tau_{\rm st}^2 \over \rho^2} \, {P^2
\over T^2} \, \langle \tau \big[{\bm\nabla}^2
\theta({\bm x})\big] \, \big[{\bm\nabla}^2
\theta({\bm y})\big] \rangle
 \label{R1}
\end{eqnarray}
(see Appendix A), where $\tau$ is the turbulent time.
In ${\bm k}$-space the correlation function
$\langle \tau \big[{\bm\nabla}^2 \theta({\bm
x})\big] \,  \big[{\bm\nabla}^2 \theta({\bm
y})\big] \rangle = \int \tilde \tau(k) \, k^4
\, \langle \theta({\bm k}) \, \theta(-{\bm k})
\rangle \, \exp \big(i {\bm k} {\bm \cdot} {\bm
R}\big) \, d{\bm k} $.
Taking into account that the correlation function
of temperature fluctuations,
$\langle \theta({\bm k})\, \theta(-{\bm k}) \rangle = \langle \theta^2
\rangle \tilde E_\theta(k) / 4 \pi k^2$, and
integrating in ${\bm k}$-space we obtain:
\begin{eqnarray}
B({\bm R}) \approx {2 \,\tau_{\rm st}^2 \, c_s^4 \,
\over 3 \, \nu} \,\left({{\bm \nabla} T \over T} \right)^2 \, {\rm Re} ,
 \label{R3}
\end{eqnarray}
where $c_s$ is the sound speed,
$\tilde E_\theta(k)=(2/3) \, k_0^{-1} \,
(k/k_0)^{-5/3}$ is the spectrum function of the
temperature fluctuations for $k_0 \leq k \leq
\ell_\eta^{-1}$, with $k_0=\ell_0^{-1}$ and
$\tilde \tau(k)=2 \tau_0 \, (k/k_0)^{-2/3}$. To
determine $\langle\theta^2 \rangle=2E_\theta$ we used the
budget equation for the temperature fluctuations:
$D E_\theta / Dt + {\rm div} \, {\bm
\Phi}_\theta = - ({\bf F} {\bm \cdot}
\bec{\nabla}) T - \varepsilon_\theta $,
that for homogeneous turbulence in a steady
state yields:
$\langle \theta^2 \rangle=- 2 \, \tau_0 \,({\bm F}
{\bf \cdot} {\bm \nabla}) T = (2/3)
\left(\ell_0 {\bm \nabla} T\right)^2$,
where $F_i = \langle u'_i \theta \rangle = -
D_{_{T}}^{(\theta)} \nabla_i T$ is the turbulent
heat flux, $D_{_{T}}^{(\theta)}=u_0 \ell_0 /3$ is
the coefficient of the turbulent diffusion of the
temperature fluctuations and the dissipation rate
of $E_\theta$ is $\varepsilon_\theta = \langle
\theta^2 \rangle /2 \tau_0$.
Equation~(\ref{R3}) implies that the correlation function
$B({\bm R})$ vanishes when the Reynolds number tends to zero.
The effect exists only in the presence of developed turbulence.

In a non-stratified turbulence $({\bm \nabla} T=0)$,
the function $B({\bm R})= 20 \sigma_{\rm v} / \tau_\eta (1 + \sigma_{\rm v})$,
where $\sigma_{\rm v} \equiv \langle ({\bm\nabla} {\bm \cdot} {\bm v}')^{2}
\rangle / \langle ({\bm\nabla} {\bm\times} {\bm v}')^{2}
\rangle$ is the degree of compressibility of the particle velocity field. For small
Stokes numbers, $\sigma_{\rm v}\approx(8/3) {\rm St}^2$, so that
$B({\bm R})= 160 \, {\rm St}^2 / 3\tau_\eta$, where we took
into account that for a Gaussian velocity field:
$\langle ({\bm\nabla} {\bm \cdot} {\bm v})^{2}
\rangle  = (80 / 3 \tau_\eta^2) \, {\rm St}^2$
and $\langle ({\bm\nabla} {\bf\times} {\bm v})^{2}
\rangle = 10 / \tau_\eta^2$ (for details see \cite{Elperin-et-al2002}).
On the other hand, for stratified turbulence $({\bm \nabla} T \not=0)$ and small
Stokes number,
\begin{eqnarray}
B({\bm R}) \approx {2 \,\tau_{\rm st}^2 \, c_s^4 \,
\over 3 \, \nu} \,\left({{\bm \nabla} T \over T} \right)^2 \, {\rm Re} = {160 \, \tilde{\rm St}^2 \over 3\tau_\eta} =  {20 \tilde \sigma_{\rm v} \over \tau_\eta},
\nonumber\\
 \label{RR3}
\end{eqnarray}
where $\tilde {\rm St} = {\rm St} \, \Gamma$, $ \tilde\sigma_{\rm v}\approx(8/3) \tilde {\rm St}^2$,
\begin{eqnarray}
\Gamma\left({\rm Re}, L_{\rm eff}/L_T \right) &=& {\rm Re}^{1/2} \,\left({L_{\rm eff} \, {\bm \nabla} T \over T}
\right) ,
 \label{RA9}
\end{eqnarray}
and variables with tilde symbols correspond to those for stratified turbulence. Here $L_{\rm eff} = c_s^2 \tau_\eta^{3/2}/9 \nu^{1/2}$ is an effective length scale,  $L_T=T/|{\bm \nabla} T|$ is the characteristic scale of the mean temperature variations.

In general case that includes both, the tangling clustering instability
and the inertial clustering instability, the parameter $\Gamma$
can be written in the following form:
\begin{eqnarray}
\Gamma\left({\rm Re}, L_{\rm eff}/L_T \right) &=& \left[1
+ {\rm Re} \,\left({L_{\rm eff} \, {\bm \nabla} T \over T}
\right)^2\right]^{1/2} ,
 \label{R9}
\end{eqnarray}
where the inertial clustering instability corresponds to the case of $\Gamma=1$.
For typical parameters
of atmospheric turbulence: (i) ${\rm Re} = 10^7$ ($u_0 = 1$ m/s, ${\ell _0}= 100$ m)
and the mean temperature gradient, $|{\bm \nabla}
T| = (0.3 - 1)$ K / 100 m, the effective length $L_{\rm eff}=23$ km, the dimensionless parameter $\Gamma = (1 - 2) \times 10^3$; (ii) ${\rm Re} = 10^6$ ($u_0 = 0.3$ m/s, ${\ell _0}= 30$ m) and the mean temperature gradient, $|{\bm \nabla}
T| = (0.3 - 1)$ K / 100 m, the effective length $L_{\rm eff}=130$ km, the dimensionless parameter $\Gamma = (1 - 4) \times 10^3$.

When the Stokes number is not small, the degree of compressibility is given by
\begin{eqnarray}
\sigma_{\rm v} ={(8/3) {\rm St}^2 \over 1 + {\rm St}^2}
 \label{RRR9}
\end{eqnarray}
(for details see \cite{Elperin-et-al2007}). Since for the stratified turbulence $B({\bm R})= 20 \tilde \sigma_{\rm v} / \tau_\eta (1 + \tilde \sigma_{\rm v})$, and $\tilde \sigma_{\rm v} =(8/3) \tilde {\rm St}^2 / (1 + \tilde {\rm St}^2)$, the function $B({\bm R})$
for the stratified turbulence and for arbitrary Stokes numbers reads:
\begin{eqnarray}
B({\bm R}) &=& {160 \, {\rm St}^2 \, \Gamma^2 \over \tau_\eta \left(3 + 11 \, {\rm St}^2 \, \Gamma^2\right)} .
\label{RR9}
\end{eqnarray}
For $\Gamma=1$ Eq.~(\ref{RR9}) describes
the inertial clustering.
When the diameter of the droplet $d \approx 1.7 \, \mu$m,
${\rm St}^2 \, \Gamma^2 \approx 3/11$.  This implies that when the diameter of the droplets is much larger than $1.7 \, \mu$m, the parameter ${\rm St}^2 \, \Gamma^2 \gg 3/11$, and the function $B({\bm R}) \sim 160/ 11 \tau_\eta$ is independent
of the Stokes number and the size of droplets.
As can be seen from Eq.~(\ref{RR9}) the effect of tangling clustering is
much stronger than the inertial clustering only for droplets smaller than 20 $\mu$m.
Remarkably, that the inertial clustering instability can be excited only if the size of droplets is larger than 20 $\mu$m (see \cite{Elperin-et-al2002}).
Analysis of the solution of Eq.~(\ref{S2}) for the two-point second-order correlation function, $\Phi(t,{\bm R})$ performed in Sect. V in \cite{Elperin-et-al2013} shows that the ratio of the minimum and maximum of the pair correlation function reads:
\begin{eqnarray}
{\Phi_{\rm min} \over \Phi_{\rm max}} = - {\pi
\over e \, \lambda} \left({{\rm Sc}^{-\lambda/2}
\over \ln {\rm Sc}}\right) ,
 \label{E9}
\end{eqnarray}
where ${\rm Sc} = \nu /{D_m}$  is the Schmidt
number, the parameter $\lambda (\tilde \sigma_{\rm v}) =
(20\tilde \sigma_{\rm v} + 1)/4(\tilde \sigma_{\rm v} + 1)$ in
Eq.~(\ref{E9}) depends on the degree of
compressibility of the particle velocity field,
$\tilde \sigma_{\rm v}$. For typical parameters
of atmospheric turbulence, parameter $\lambda$ varies
in the range from $0.5$ to $2.5$.

As follows from Eq.~(\ref{RR9}),
the temperature fluctuations, which are caused by the tangling
of the mean temperature gradient, ${\bm \nabla} T$,
by the fluid velocity fluctuations ${\bm u}'$, strongly contribute to
the function $B({\bm R})$ and the growth rate of
the tangling clustering instability in the
temperature-stratified turbulence.
The mechanism of coupling related to the tangling
of the gradient of the mean temperature gradient
is quite robust. The tangling
is not sensitive to the exponent of the
energy spectrum of the background turbulence.
Anisotropy effects do not introduce new physics
in the clustering process because the main
contribution to the tangling clustering
instability is at the Kolmogorov (viscous) scale
of turbulent motions, where turbulence can be
considered as nearly isotropic, while anisotropy
effects can be essential in the vicinity of the
maximum scales of the turbulent motions.

Equation~(\ref{R9}) shows that
the tangling clustering instability can be much more
effective than the inertial clustering
instability which is excited in a non-stratified
turbulence \cite{Elperin-et-al2002}. In both
instabilities, the particle clustering is
determined by the two-point correlation function
of the Laplacian of air pressure fluctuations.
However, in case of non-stratified turbulence the
pressure fluctuations are of the order of  $\rho{\bm u}'^2$, while in case of the
temperature-stratified turbulence there are
additional pressure fluctuations caused by
temperature fluctuations, $p' \propto P\left(\theta /T \right)
\propto (P/T) \ell_0 |{\bm \nabla} T|$,
where we took into account that the root mean square of the tangling temperature fluctuations
$\theta \sim \ell_0 |{\bm \nabla} T|$ (see \cite{TL73}). Consequently, the ratio of the two-point
correlation functions of the Laplacian of air
pressure fluctuations in stratified and
non-stratified flows is proportional to
\begin{eqnarray}
{B_{\rm tangling} \over B_{\rm isothermal}}
\sim \left({\tilde n k_B T \over \rho {\bm u}^2} \right)^2 {\ell_0^2 \over L_T^2} ,
 \label{PR9}
\end{eqnarray}
which is a large parameter because
the thermal energy density $\tilde n k_B T$ is much larger
than the turbulent kinetic energy
$\rho {\bm u}'^2$, where $\tilde n$ is the number density of molecules.
Here we used the equation of state for the ideal gas $P=\tilde n k_B T$.

Due to inertia effects droplets accumulate in the regions
with the increased pressure of the air flow.
The effect of increased pressure fluctuations
in a temperature stratified turbulence is more pronounced in small
scales because the function $B({\bm R})$ is determined by
the two-point correlation function of the Laplacian of pressure fluctuations.
The tangling clustering is
also enhanced by the effect of turbulent thermal
diffusion \cite{Elperin-et-al1996a} that causes
non-diffusive streaming of particles in the
direction of a heat flux and accumulation of
particles in the regions with the minimum mean
temperature of the air flow. Temperature
fluctuations in the stratified turbulence produce
pressure fluctuations and cause particle
clustering due to the tangling clustering
instability with the growth rate, that is
by a factor ${\rm Re} \left(L_{\rm eff}/L_T \right)^2$
larger than the growth rate of the inertial
clustering instability [see Eq.~(\ref{R9})].
For large Reynolds numbers the tangling mechanism
is universal and weakly dependent on the origin of
turbulence.

\subsection{Growth rate of the instability}

To illustrate the tangling clustering
instability we use the standard dependence of
the droplet evaporation time on their diameter and the relative humidity
[see Eq.~(\ref{Eq8}) below].
Figure 1 shows the growth rate of
the instability (measured in the inverse
turbulent Kolmogorov time scale units,
$\tau_\eta^{-1}$) versus the droplet diameter $d$
(measured in $\mu$m) for different values of
relative humidity, $\phi$, i.e., for very low
humidity (45$\%$ and 90$\%$) and for very high
humidity (99$\%$ and 100$\%$). Inspection of
Fig.~1 shows that the threshold for the tangling
clustering instability based on the size of the droplets is
$d_{\rm th} = 0.7 \mu m$. The instability is excited when
$d > d_{\rm th}$, and there is a sharp maximum of the
growth rate of the instability at $d = 1.75 \, \mu$m  if the
relative humidity is close to saturation,
99$\%$ and 100$\%$. This explains the fast
growth of the droplets having the initial
diameter of the order of 1.75 $\mu$m caused by the
tangling clustering instability. For lower values
of the relative humidity the threshold for the tangling
clustering instability increases and the droplet growth rate
sharply decreases. For $d \geq 5 \mu$m and for very high
humidity (99$\%$ and 100$\%$) the growth rate
of the tangling clustering instability is constant and
independent of the droplet size.

\begin{figure}
\vspace*{1mm} \centering
\includegraphics[width=10.5cm]{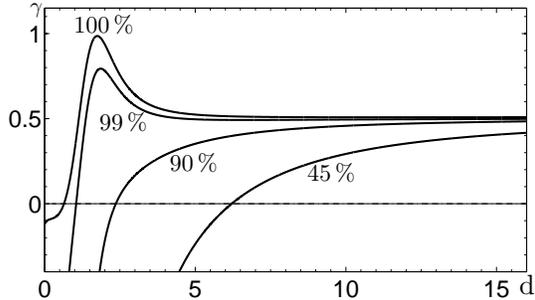}
\caption{\label{Fig1} Growth rate $\gamma$ of the
tangling clustering instability (measured in
units $1/\tau_\eta$) versus the droplet diameter
$d$ (measured in $\mu$m) for $\Gamma  = 10^3$
and ${\rm Sc} = \nu /{D_m} = 5 \times
10^5 d (\mu$m).}
\end{figure}

The exponential growth of droplet number density
inside the cluster is saturated by nonlinear
effects. The droplet number density inside the
cluster can be constraint by depletion of
particles in the surrounding air flow caused by
their accumulation inside the cluster. Another
effect that inhibits the growth of the droplet
number density inside the cluster is related to a
strong momentum coupling of particles and
turbulent air flow when the mass loading
parameter $m_{\rm dr} n_{\rm max}/\rho  \approx
0.5$.

It can be shown \cite{Elperin-et-al2013}
that the maximum increase of particle number
density inside the cluster, $n_{\rm max}/ N$,
caused by the first effect is:
\begin{eqnarray}
{n_{\rm max} \over N} = \left(1 + {e \, \lambda
\over \pi} \, {\rm Sc}^{\lambda/2} \, \ln {\rm
Sc}\right)^{1/2} .
 \label{Eq4}
\end{eqnarray}
In this analysis small yet finite
molecular diffusion $D_m$ has been taken into account.
In the limit $D_m=0$ Eq.~(\ref{Eq4}) is not valid.
For instance,
the Schmidt number for droplets in the atmospheric flow
is ${\rm Sc}= 5 \times 10^5 \, d (\mu$m).

\begin{figure}
\vspace*{1mm} \centering
\includegraphics[width=10.5cm]{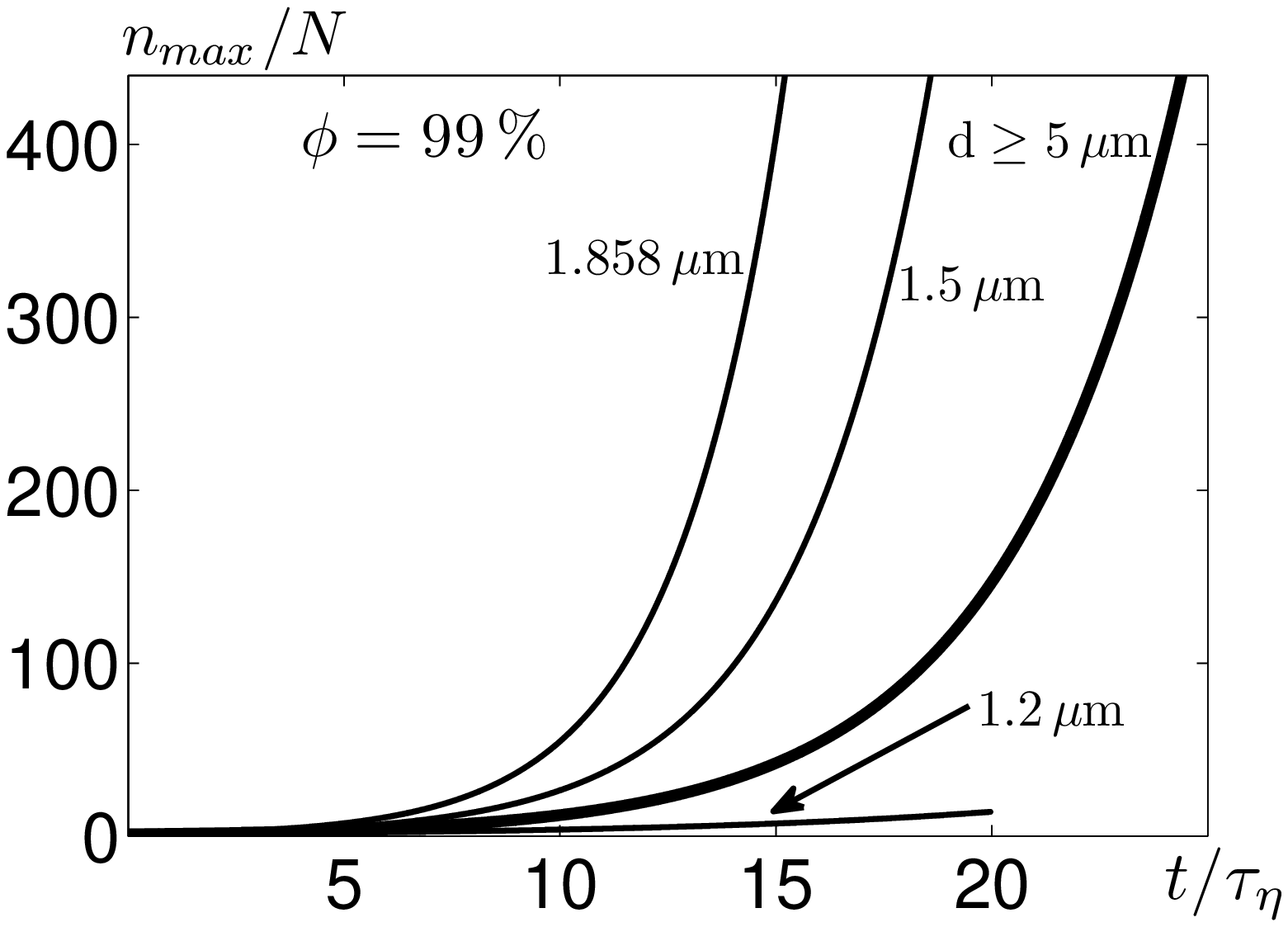}
\includegraphics[width=10.5cm]{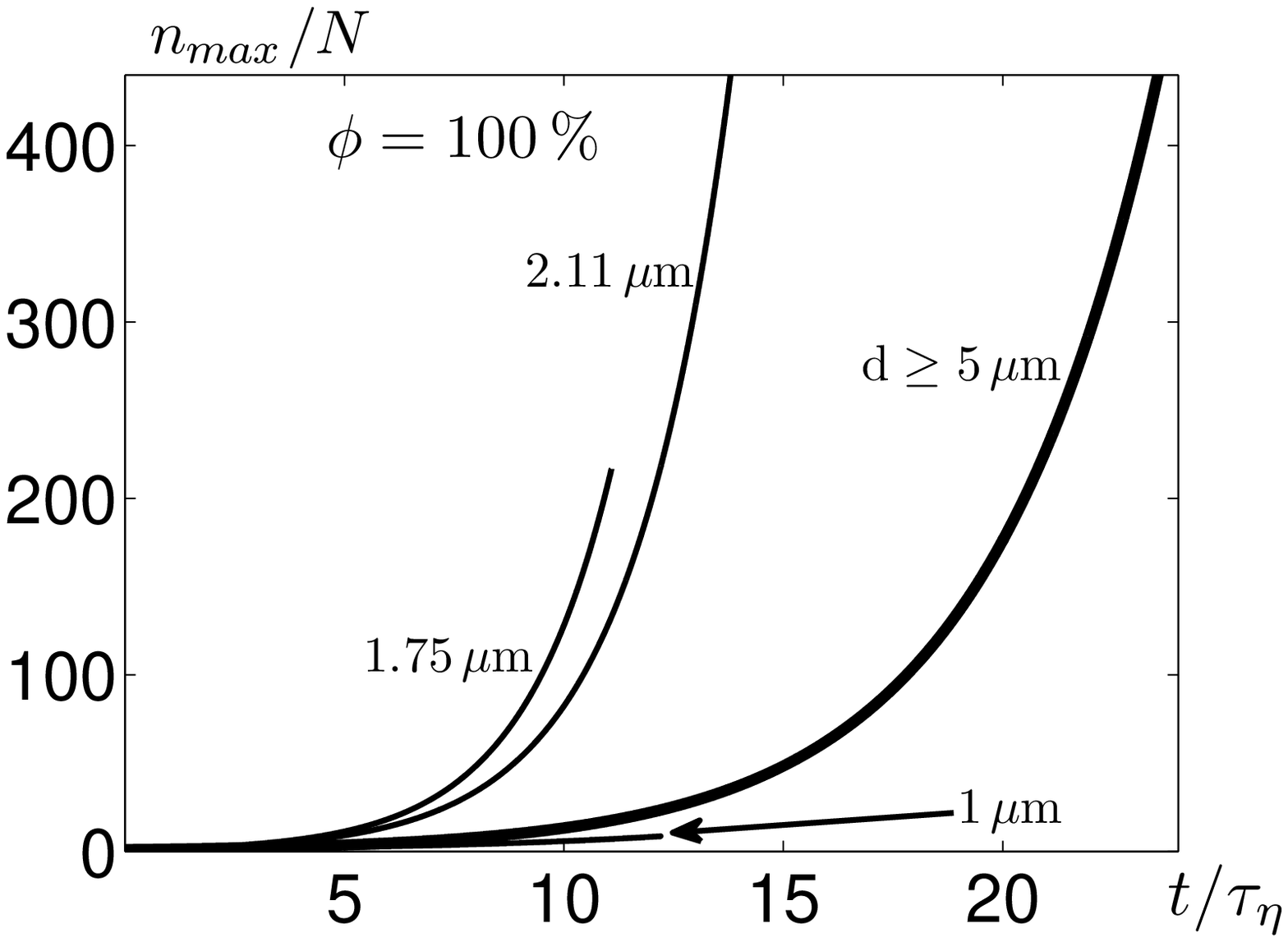}
\caption{\label{Fig2} The time evolution of
droplet concentration $n_{\rm max}/N$ inside the
cluster for droplets of different diameter for
relative humidity 99 $\%$ (upper panel) and 100
$\%$ (bottom panel); $\Gamma  = 10^3$ and Sc $= 5
\times 10^5 \, d$. Thick solid line corresponds to droplets with $d \geq 5 \mu$m
(for which the growth rate of the tangling clustering instability is independent
of the droplet size).}
\end{figure}

Let us estimate the limitation caused by strong momentum coupling of
particles and turbulent air flow. Assuming as
total cloud water content $\bar \rho_{\rm dr}
\equiv m_{\rm dr} n_{\rm max} = 1.5$ g/m$^3$ and
taking into account that the air density in the
atmosphere $\rho \approx 1.3 \times 10^3$
g/m${^3}$ , we obtain:
$n_{\rm max}/N < 0.5\,\rho /\bar\rho_{\rm dr}
\approx 470$.

Figure 2 shows the increase of
droplets concentration $n_{\rm max}/N$  inside
the cluster during development of the tangling
clustering instability for the relative humidity
$99~\%$ and $100~\%$ and for different droplet sizes. The growth rate of the
instability is of the order of 10 inverse
Kolmogorov time-scales, where $\tau_\eta \approx
0.1$ s.
In this study we consider only droplet
clustering, but not aerosol dynamics. Clearly,
the clustering of aerosols is similar to that of
droplets for humidity of 100 $\%$, i.e. without
evaporation.

Strong increase of clustering in a temperature stratified fluid
in comparison with the inertial clustering has been confirmed
in the laboratory experiments \cite{Eidelman-et-al2010}.
The experimental study of the particle
clustering compared the two-point correlation functions
for both inertial and tangling clustering measured
with sub-Kolmogorov scale resolution.
The experimental parameters were: the r.m.s. velocity
$u_0 = 12$ cm/s,
the integral (maximum) scale of turbulence $\ell_0 = 3.2$ cm,
the Reynolds numbers ${\rm Re} =250$,
the Kolmogorov length scale $\ell_\eta
= 510 \, \mu$m and the Kolmogorov time scale
$\tau_\eta = 1.7 \times 10^{-2}$ s.
The Stokes time for the particles with the diameter $d = 10
\, \mu$m is $\tau_{\rm st} = 10^{-3}$ s, the Stokes number ${\rm St}
= 5.9 \times 10^{-2}$, the coefficient of
molecular diffusion $D_m = 1.4 \times 10^{-8}$ cm$^2$ /s
and the Peclet number Pe$=u_0 \, \ell_0 / D_m =3 \times 10^9$.
These experiments demonstrated that the two-point
correlation function of the particle number density
fluctuations for the tangling clustering in
temperature stratified turbulence is by one order of magnitude
larger than that for the inertial clustering
in isothermal turbulence \cite{Eidelman-et-al2010}.
This is consistent with the efficiency of the tangling clustering
being proportional to ${\rm Re}^{1/2}$ [see Eq.~(\ref{R9})].
In these laboratory experiments ${\rm Re}^{1/2}\sim 15$.
Since in the atmospheric turbulence the Reynolds number is $10^7$
(${\rm Re}^{1/2} \sim 3 \times 10^3$)
it is plausible to suggest that for atmospheric conditions the effect of the
tangling clustering will be more pronounced.

\section{Detection of droplet clustering in atmospheric cloud measurements}

In this Section we present a short discussion of the existing cloud measurements
and their relation to the droplet clustering.
Aircraft mounted with forward-scattering spectrometer probe \cite{B92} was used
for study of 1 cm droplet concentrations in cumulus clouds. In these experiments
over 50 cloud passes have been done.
The statistical analysis in \cite{B92} has shown
significant deviations from the Poisson distribution,
which characterizes a random homogeneous spatial distribution
of the droplets. These findings were interpreted in \cite{B92} as appearance of small-scale
(about 1 cm) droplet  clusters in cumulus clouds.

The droplet clusters have been detected in \cite{PK01} by
analyzing the measurements obtained in situ in 57 clouds
by the Fast Forward-Scattering Spectrometer Probe (FSSP).
This finding is the direct evidence of turbulence-inertia
impact on droplet motion in clouds. Dissipation rate of
the turbulent kinetic energy in clouds was varied
in the atmospheric measurements \cite{PK01} from $10^{-4}$ to $2.3
\times 10^{-2}$ m$^{2}$ s$^{-3}$.
The rms of small-scale droplet concentration fluctuations
was estimated to be about 31 \% of
the mean values of droplet concentration both over the whole cloud
and in a more homogeneous adiabatic core.
The power spectrum shows that fluctuations with spatial scales
within the 0.5 - 5 cm range contain over 80 \% of
the energy of small-scale fluctuations \cite{PK01}.
An increase in turbulence intensity and droplet inertia
result in an increase of the droplet concentration fluctuations.

In other experiments \cite{KS01} the droplet positions have been measured with
the Meteo-France Fast Forward Scattering Spectrometer Probe.
The cloud droplet data have been collected during a single traverse by
the Meteo-France Merlin IV research aircraft through a cumulus cloud encountered
during the Small Cumulus Microphysics Study.
The energy dissipation rate was of the order of
$10^{-4}$ m$^{2}$ s$^{-3}$.
The collected data in \cite{KS01} reveal droplet clustering even in cumulus cloud
cores free of entrained ambient air.
The pair correlation function has been obtained
in \cite{KS01} for droplets in a high Reynolds number turbulent flow.
The super-Poissonian variances which were detected in these homogeneous core data,
were viewed in \cite{KS01} as conclusive evidence of clustering.
It was shown in \cite{SKL02} by using the correlation-fluctuation theorem and the Wiener-Khinchin theorem, that the pair-correlation function is ideal for quantifying
droplet clustering because it contains no scale memory and because of its quantitative
link to the Poisson process.

Simultaneous observations of cloud droplet spatial statistics, cloud droplet size distribution and cloud turbulence were made in \cite{LS07} during several cloud passages, including cumulus clouds and a stratus cloud. The measurements were conducted using the Airborne Cloud-Turbulence Observation System (ACTOS) which was suspended
from a tethered balloon. The ACTOS instrumental payload was equipped with sensors to measure
the three-dimensional wind velocity, static air temperature,
and humidity with a sampling frequency of at least 100 Hz.
The wind velocity was measured by an ultrasonic anemometer.
Cloud droplet number density and droplet size distribution
were obtained from measurements with the M-Fast-FSSP, which records sizes and arrival
times of individual droplets.
The primary finding of the study in \cite{LS07} by the determining of the droplet
pair correlation function (with the spatial resolution about $100 - 200 \mu$m)
is the indication of the droplet
clustering even for small Stokes numbers (smaller than $10^{-2})$ and also in weakly turbulent clouds (with the dissipation rate of the turbulent kinetic energy that is smaller than $10^{-2}$ m$^{2}$ s$^{-3}$). For three analyzed
cases, two horizontal passages through cumulus clouds and vertical
profiles through a stratus cloud, the regions where
droplets are clustered at sub-cm scales, have been found in \cite{LS07}.

All these studies for the most part show very modest clustering at Kolmogorov
separations and below. Note, however, that the FSSP measures droplets
along a narrow almost 1-D horizontal path through a
cloud volume, which means that a long sample is necessary
to construct a reasonable spectrum \cite{Devenish-et-al2012}. The interpretation of
the results remains controversial because deviations from
Poisson distributions could be possible due to instrumental
artefacts and the necessarily limited samples that are
obtained from aircraft measurements, which inevitably
compromise the assumption of the statistical homogeneity of
the sample \cite{Devenish-et-al2012}.
Unfortunately, the detailed measurements during all these atmospheric cloud experiments
of the spatial temperature distributions and
of the vertical and horizontal heat fluxes in clouds have not been
presented in the papers discussing the droplet clustering. Consequently, we cannot
make any conclusions about tangling clustering in these experiments.

It must be emphasized that the pair correlation function $\Phi({\bm R})$
for clustered droplet population measured in \cite{LS07} (see Fig.~1 in \cite{LS07})
agrees with the pair correlation function determined analytically
in our previous study (see Eqs.~(43) and~(47) in Ref.~\cite{Elperin-et-al2013}).
The discrepancy occurs only in the scales smaller than a Kolmogorov scale
(that is $\ell_\eta=2$ mm in \cite{LS07}). Our theory predicts that
the pair correlation function vanishes in the vicinity of $\ell_\eta$
in agreement with the atmospheric experiments (reported in \cite{LS07})
as well as with our laboratory experiments (see Ref.~\cite{Eidelman-et-al2010}).
However, the pair correlation function $\Phi({\bm R})$ according to our theory
(see Ref.~\cite{Elperin-et-al2013}) sharply increases at smaller scales.
The ratio of the minimum and maximum of the pair correlation function,
$\Phi_{\rm min} / \Phi_{\rm max}$ is given by Eq.~(\ref{E9}) of the present paper
(or Eq.~(62) in Ref.~\cite{Elperin-et-al2013}). Using the value of $\Phi_{\rm min}=-0.05$
measured in \cite{LS07} and the parameters of turbulence and droplets
for the atmospheric experiments in \cite{LS07} we find that
${\rm Sc}\equiv{\rm Pe}/{\rm Re}=3 \times 10^4$ and the ratio $\Phi_{\rm max}^{1/2}/N$
is of the order of 500, where ${\rm Pe}$ is
the droplet Peclet number. This value of the ratio $\Phi_{\rm max}^{1/2}/N$
agrees with the estimated value 470 obtained in \cite{Elperin-et-al2013}.
To determine the pair correlation function in the scales much smaller than the
Kolmogorov scale, the spatial resolution of the atmospheric measurements reported in \cite{LS07} should be improved by a factor of 10 at least.  Conducting measurements in these scales may require to abandon the Taylor hypothesis
and to employ the Particle Image Velocimetry or holographic techniques.
In this case the radial distribution function (RDF),
$G({\bf R}) =\langle n(t,{\bf x}) n(t,{\bf y}) \rangle / N(t,{\bf x}) N(t,{\bf y})$
can be determined from two-dimensional images of a field of $M$ droplets by binning the droplet pairs according to their separation distance, so that the function $G({\bf R})$ is determined as follows:
\begin{eqnarray}
G({\bf R}) \approx {N_{\Delta S}^{(p)} / \Delta S \over N_S^{(p)} / S} ,
\label{E2}
\end{eqnarray}
where $\Delta S = \pi [(R + \Delta R/2)^2 - (R - \Delta R/2)^2]$ is the area of the annular domain located between $R \pm \Delta R/2$, $\, S$ is the area of the part of the image with the radius $R_{\rm max}$ that is used in data processing in order to exclude the edge effects.
The measured radial distribution function allows to determine the two-point correlation function of the droplet number density, $\Phi(t,{\bf R})= N^2 \, [G(t,{\bf R}) - 1]$.

In order to attain a high spatial resolution the following method should be used: (i) to determine the response function for the CCD camera by analyzing the light intensity distribution in the image for single droplet located at the center of the pixel in the form of the Gaussian distribution; (ii) segmentation of the image using a threshold technique; (iii) identification of droplets locations in the segments by least-square fitting of the recorded light intensity distribution and the light intensity distribution caused by superposition of the Gaussian distributions at the droplet locations (for details see Ref.~\cite{Eidelman-et-al2010}).

To detect the tangling clustering in the atmospheric clouds,
the measurements of the spatial temperature distributions in clouds,
as well as the fluid velocity measurements should be conducted in addition to
the measurements of RDF of droplets. In particular, it is important to measure
the vertical and horizontal heat fluxes, $\langle {\bf u} \, \theta \rangle$ and
two-point correlation functions of the temperature fluctuations, $\langle \theta({\bf x}) \, \theta({\bf y}) \rangle$, in clouds. This allows to determine the rate of tangling clustering $B({\bf R})$ [see Eqs.~(\ref{R1}) and~(\ref{RR9})].
In addition, measurements of two-point non-instantaneous correlation functions of fluid velocity allow to determine integral scale of turbulence and turbulent time scales.
Measurements of turbulent fluxes of droplets $\langle {\bf u} \, n' \rangle$ in combination
with the measurements of spatial distributions of droplets allow to determine turbulent diffusion coefficients of droplets.

\section{Collision kernel and droplet coagulation}

In this Section we consider droplet coagulation
and apply the theory of the tangling clustering instability
to explain acceleration of raindrops formation in warm clouds.
The warm clouds often exist in the region of atmospheric turbulent convection
with coherent structures (cloud ``cells'' in shear-free convection
and cloud ``streets'' in sheared convection, see e.g., \cite{EB93,AZ96}).
The vertical large-scale temperature gradient is small inside
the large-scale circulation (coherent structures) in a small-scale turbulent
convection.
However, the horizontal large-scale temperature gradient
inside the circulations is not small. Atmospheric observations
showed that this gradient is about 1 K per 100 m \cite{WH92}.
Similar results were reported in laboratory experiments where the horizontal
large-scale temperature gradient inside the large-scale circulation was
0.6 K per 1 cm,  while the vertical large-scale temperature gradient was 0.05 K per 1 cm  \cite{BEKR09}. This magnitude of the horizontal temperature gradient
is sufficient for the generation of strong temperature fluctuations in the
stratified turbulence by tangling mechanism.

The initial stage of cloud droplets formation
involves condensation of water vapor on cloud
condensation nuclei (CCN) and formation of small
micron size droplets. In the present study we
show that the tangling clustering instability
strongly enhances the growth rate of cloud droplets at
both stages: at the first stage when droplets
grow from the micron size to 10
$\mu$m droplets and at the next stage from 10 to
50 $\mu$m radius droplets.

\subsection{Smoluchowski coagulation equation}

Subsequent evolution and growth of small droplets
due to collision-coalescence depend on the
interplay between their collision time and
evaporation time, in particular because of water
vapor depletion. The collision time of small
droplets can be determined using the Smoluchowski
coagulation equation (see, e.g.,
\cite{Seinfeld-and-Pandis2006}, Chapter 13):
\begin{eqnarray}
&&{\partial {\tilde n}(d) \over \partial t} + {\rm div} \left({\tilde n}
\, {\bm v} \right) - D_m \Delta {\tilde n} + {{\tilde n} \over \tau
_{ev}}
\nonumber\\
&& = {1 \over 2} \int_0^d \, K(\hat d,x) \,
{\tilde n}(\hat d) \, {\tilde n}(x) \,dx - \int_0^\infty K(d,x) \,
{\tilde n}(x) \, {\tilde n}(d) \,dx  ,
\nonumber\\
 \label{Eq5}
\end{eqnarray}
where $\hat d = \left(d^3 - x^3\right)^{1/3}$,
${\tilde n}(d)$ is the droplet size distribution,
$n= \int{\tilde n}(x) \, dx$
is number density of droplets, and  $K(d,x)$ is the coagulation
kernel that describes coagulation rate of
droplets of the diameter $d$ and droplets of the
diameter $x$. In the present study we use the
coagulation kernel $K(d,x)$ as a sum of the
Brownian coagulation kernel (see Table 13.1, p.
600 in \cite{Seinfeld-and-Pandis2006}) and the
gravitational coagulation kernel (see Eq.
(13.A.4), p. 615 in
\cite{Seinfeld-and-Pandis2006}).

Averaging Eq.~(\ref{Eq5}) over the statistics of particle
turbulent velocity field, estimating integrals in
Eq.~(\ref{Eq5}), using the mean-value theorem, and taking
into account that  $\left\langle {\tilde n}(d)\,{\tilde n}(d_1)
\right\rangle$ is calculated in the same point,
so that $\left\langle {\tilde n}(d)\,{\tilde n}(d_1) \right\rangle
\le {\tilde n}_{\rm max}(d)\,{\tilde n}_{\rm max}(d_1) = C(d,d_1)
\, {\tilde N}(d) \,{\tilde N}(d_1)$, we obtain the following
equation for the mean droplet size distribution
${\tilde N}(d)$:
\begin{eqnarray}
{\partial {\tilde N}(d) \over \partial t} + {\rm{div}}
\left({\tilde N}\, {\bm V}_{\rm dr} + \left\langle {\tilde n}' \, {\bm u}
\right\rangle \right) &=&  - {{\tilde N} \over \tau_{\rm
ev}(d)} - {{\tilde N} \over \tau_{\rm eff}^{\rm st} (d)}
\nonumber\\
&& + D_T \, \Delta {\tilde N},
 \label{Eq6}
\end{eqnarray}
where $C(d,d_1) = {\tilde n}_{\rm max}(d) \, {\tilde n}_{\rm
max}(d_1) / {\tilde N}(d) \, {\tilde N}(d_1)$, $D_T(d)$  is the
turbulent diffusion coefficient and
\begin{eqnarray}
\tau_{\rm eff}^{\rm st}(d) &=& {1 \over {\tilde N}(d) \,
K(d,d_1) \, C(d,d_1)}
\nonumber\\
&& \quad > {1 \over {\tilde N}(d) \,
K(d,d_1)} \left({{\tilde n}_{\rm max}(d) \over {\tilde N}(d)}
\right)^{-2} . \label{Eq7}
\end{eqnarray}
Notably, the collision term ${\tilde N}(d)/\tau_{\rm
eff}^{\rm st}(d)$ in Eq.~(\ref{Eq6}) is similar to the
droplet evaporation term.
The coefficient of molecular
diffusion of droplets having the diameter $d$ in the atmosphere is
$D_m = 2 \times 10^{-7} / \, d (\mu m)$
cm$^2$ s$^{-1}$,
while the turbulent diffusion coefficient
$D_T= u_0 \ell_0 / 3 = 3 \times 10^{5}$
cm$^2$ s$^{-1}$, where turbulent velocity $u_0$ at the
integral turbulent scale ${\ell _0}= 100$ m is $u_0 = 1$ m/s.
Therefore, the coefficient of molecular
diffusion of droplets is much smaller than the turbulent diffusion coefficient.

\begin{figure}
\vspace*{1mm} \centering
\includegraphics[width=10cm]{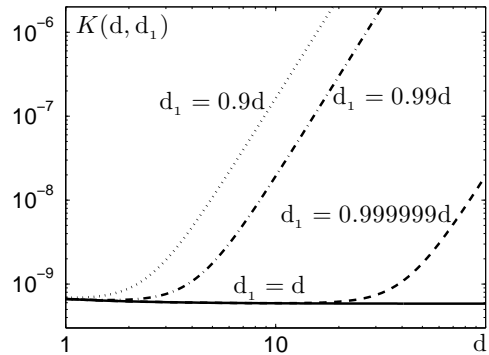}
\caption{\label{Fig3} Sum of Brownian and
gravitational coagulation kernels versus the
droplet diameter $d$ in microns for two droplets
having different diameters: $d = 0.9 d_1$  -
dotted line; $d = 0.99 d_1$  - dashed-dotted
line; $d - d_1 = 10^{-6}d$   - dashed line. Solid
line is Brownian coagulation kernel $K(d,d_1)$
[measured in cm$^3$/s] for two droplets having
equal diameters $d = d_1$. The diameter $d$ of droplets is measured in $\mu$m.}
\end{figure}

\subsection{Effective collision-coalescence time}

Now we can estimate the droplet collision time
and compare it with the evaporation time of
droplets having different sizes. The most
interesting case is the growth of droplets when
the relative humidity is only slightly less
100$\% $ and the evaporation of droplets competes
with their coagulation. Figure~3 shows the
numerical values of the sum of the Brownian and
gravitational coagulation kernels versus droplet
diameter $d$ when droplets have the same or
different sizes \cite{Seinfeld-and-Pandis2006}.
Inspection of Fig. 3 shows that the collision
kernel varies slightly when $d < 2$ $\mu$m, and
it increases by one order of magnitude for $d =
5$ $\mu$m, while for $d > 5$ $\mu$m the collision
kernel can increase by three orders of magnitude
depending on the difference in size of colliding
droplets ($d$  and $d_1$). However, the effect of
this increase on the droplet collision rate is
much smaller than the increase of droplet
collision rate due to increase of the droplet
number density caused by the tangling clustering
instability that is up to five orders of magnitude.

Dynamics of the raindrops evolution and their
growth depend on the interplay between the
characteristic times of droplet collisions
resulting in droplet coagulation and the time of
droplet evaporation. The characteristic times of
vapor diffusion and thermal relaxation in the
gaseous phase in the vicinity of a droplet can be
estimated as  $\tau_{\rm dif} \propto d^2/D_{\rm
v}$, and $\tau_{\rm th} \propto d^2/\chi$, where
$D_{\rm v} = 0.216$ cm$^2$ s$^{-1}$ is
coefficient of binary diffusion of water vapor in
air and  $\chi  = 0.185$ cm$^2$ s$^{-1}$ is
thermal diffusivity of air
\cite{Seinfeld-and-Pandis2006}. Since these
characteristic times are much smaller than the
time of droplet evaporation or growth, the
evaporation/growth of cloud droplets is
determined by stationary vapor diffusion. In this
case the characteristic time of the decrease of
droplet radius due to evaporation can be
estimated using the coupled analytical model of
the evaporation/growth rates of droplets (see Ref.~\cite{Nadykto-et-al2003}).
For the ambient air temperature $T_a = 274$ K, this model yields the
following expression for the evaporation time:
\begin{eqnarray}
\tau _{\rm ev} = 0.5 \times 10^{-3} {d^2 \over 1
- \phi},
\label{Eq8}
\end{eqnarray}
where the droplet diameter is measured in microns
and time is given in seconds. The calculated
evaporation times versus droplet radius for
relative humidity  $\phi  = 99 \%$ and $\phi  =
99.99 \%$ together with the effective
collision-coalescence time within the cluster are
shown in Fig.~4. To determine the effective
collision-coalescence time we have assumed that a
total cloud water content of mean droplets mass
density is about $\bar\rho_{\rm dr} = 1.5\,{\rm
g/m}^3$ , which corresponds to the typical mean
number density of 10 $\mu$m droplets,
$N \approx 2$ ${\rm cm}^{-3}$, while for 2 $\mu$m
droplets it is about $N \approx 2 \times 10^2$
cm$^{-3}$.

\begin{figure}
\vspace*{1mm} \centering
\includegraphics[width=8cm]{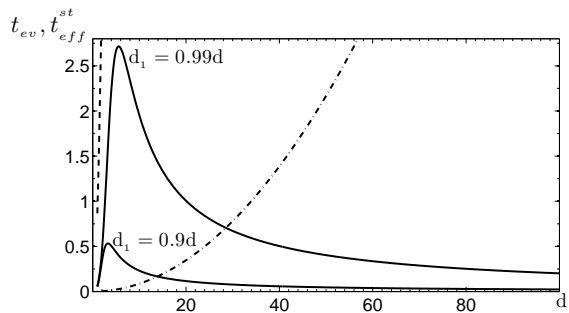}
\caption{\label{Fig4} Evaporation times versus
droplet diameter for relative humidity $\phi  =
99\% \,$ (dashed-dotted line) and  $\phi = 99.99
\% $  (dashed line) and effective
collision-coalescence time (solid lines for
different $d - d_1$).
The diameter $d$ of droplets is measured in $\mu$m
and time is measured in minutes.}
\end{figure}

In the absence of tangling clustering
instability, for the ambient number density of
the micron size droplets having the mean number
density $N \approx 10^2$ cm$^{-3}$ , the
collision-coalescence time is of the order of
$\tau_{\rm eff}^{\rm st}(d = 2\,\mu$m) $\approx
({\tilde N} \, K)^{-1} \approx 10^7$ s, and for droplets with
diameter $d = 10\,\mu$m  and the mean number
density  $N(d = 10\,\mu$m) $\approx 1$
cm$^{-3}$, the collision-coalescence time is
$\tau_{\rm eff}^{\rm st}(d = 10 \,\mu{\rm
m}) > 10^7$ s. These values are too
large to account for the collision-coalescence
growth of cloud droplets since the droplet
evaporation time is much less than their
collision time. The latter conclusion implies
that small micron-size and submicron-size
droplets are either in equilibrium or grow very
slowly due to condensation of supersaturated
water vapor. In these calculations we have taken
into account kinetic corrections to
submicron-size droplet evaporation time using the
flux-matching approach suggested in \cite{Lushnikov-and-Kulmala2004}.

The situation drastically changes in the presence
of the tangling clustering instability. In this
case the droplet collision time inside the
clusters, which are formed due to the tangling
clustering instability,  decreases by the large
factor, $[n_{\rm max} /N]^2 \sim 10^5$.
Indeed, the number density of droplets inside the
cluster sharply increases and their effective
collision time dramatically decreases:
\begin{eqnarray}
\tau_{\rm eff}^{\rm st} = {1 \over
{\tilde N}(d) \, K(d,d_1)} \left({n_{\rm max}\over
N}\right)^{-2}.
 \label{Eq9}
\end{eqnarray}
Using the numerical values of the coagulation
kernel showed in Fig.~3 we can estimate the
effective collision-coalescence time inside the
cluster. The equilibrium between the effective
droplet collision-coalescence and droplet
evaporation depends on the value of the relative
humidity  $\phi$ and the temperature of the
ambient air. The calculated effective collision
times inside the cluster for two typical values
of the relative humidity for $T = 274$ K  versus
the droplet diameter are shown by solid lines in
Fig.~4.

Using data shown in Fig.~4, we estimate the time
of growth of droplets by cascade of successive
collisions of droplets having close diameters
(with diameters ratios $d_1 / d = 1.1$ or $d_1 /
d = 1.01$). In the calculations we take into
account that after each collision droplet
diameter increases, and effective droplet
collision time changes non-monotonically as shown
in Fig.~4. For $d_1 / d = 1.1$  the time of
droplet size growth from 1 $\mu$m to 10 $\mu$m
diameter is about 3 minutes, while for $d_1 / d =
1.01$  this time is approximately  11 minutes.
The time required for further droplet size growth
from 10 $\mu$m to  (50 - 60) $\mu$m diameter
droplets is about 1 min for $d_1 / d = 1.1$ and
5.5 minutes for colliding droplets diameters
ratio  $d_1 / d = 1.01$. It should be noted that
real droplet size growth time can be shorter due
to direct enhancement of droplet collision kernel
by turbulence (see \cite{Khain-et-al2007}
and references therein). Since droplet
collisional growth time is smaller for droplets
with larger diameters ratios, the estimated
droplet growth time can be considered as a fairly
reasonable estimate of the time required for
droplet growth. The total time required for
collisional growth of droplets having diameter 1
$\mu m$   to droplets having diameter about 50
$\mu m$  is of the order of 15 minutes that is
close to the observed $(15 - 20)$  minutes
required for formation of rain droplets.

\section{Conclusions}

New effect of the tangling clustering instability
of small droplets in turbulent temperature
stratified atmosphere results in the formation of
clusters with drastically increased droplet
number density and, correspondingly, sharply
increased rate of their collision-coalescence.
Without the tangling clustering instability, the
droplets collision-coalescence time is much
larger than the characteristic time of the
droplet evaporation. Consequently, in the absence
of tangling clustering instability droplets do
not grow due to collision-coalescence, and rain
droplets are not formed. On the contrary, in the
presence of tangling clustering instability the
effective collision-coalescence time inside the
clusters strongly decreases by the factor
$[n_{\rm max} /N]^2 \sim 10^5$. As the result,
droplets within the cluster coalesce and grow
forming large rain droplets. The growth time of
droplets from the initial size of 1 $\mu$m up to
the size of about 50 $\mu$m is $15 - 20$ minutes.

{\it In summary}, we can conclude that the effect of the tangling
clustering instability provides a convincing
explanation of the observed fast growth of cloud
droplets.

\bigskip

\begin{acknowledgements}
This work has been supported by the Israel
Science Foundation governed by the Israeli
Academy of Sciences (grant No. 1037/11);
the Research Council of Norway
under the FRINATEK (grant  No. 231444),
and grant of Russian Ministry of
Science and Education (Program 1.5/XX, contract
No. 8648). NK and IR thank NORDITA for hospitality and support during their visits.
The part of this work was completed while participating at
the NORDITA program on ``Dynamics of Particles in Flows: Fundamentals and Applications''.

\end{acknowledgements}

\appendix
\section{Derivation of the function $B({\bm R})$}

Let us determine the functions $B({\bm R})$:
\begin{widetext}
\begin{eqnarray}
B({\bm R}) &\approx& {2 \tau_{\rm st}^2 \over \rho^2} \, \langle \tau \big[\bec{\nabla}^2 p'({\bm x}) \big]\, \bec{\nabla}^2 p'({\bm y}) \rangle \approx
{2 \tau_{\rm st}^2 \over \rho^2} \, \Big[ {P^2 \over T^2} \, \langle \tau \big[\bec{\nabla}^2 \theta({\bm x})\big] \, \bec{\nabla}^2 \theta({\bm y}) \rangle
+ {P^2 \over \rho^2} \, \langle \tau \big[\bec{\nabla}^2 \rho'({\bm x})\big] \, \bec{\nabla}^2 \rho'({\bm y}) \rangle
\nonumber\\
&& + {P^2 \over \rho \, T} \, \Big(\langle \tau \big[\bec{\nabla}^2 \rho'({\bm x})\big] \, \bec{\nabla}^2 \theta({\bm y}) \rangle
+ \langle \tau \big[\bec{\nabla}^2 \theta({\bm x})\big] \, \bec{\nabla}^2 \rho'({\bm y}) \rangle \Big) \Big] \;,
\label{AL5}
\end{eqnarray}
\end{widetext}
\noindent
where $\bec{\nabla}^2 p'({\bm x}) = \big[\bec\nabla^{({\bm x})} \big]^2 p'({\bm x})$,
and $\rho'$ are the fluid density fluctuations.
In derivation of this equation we used the relationship
\begin{eqnarray}
{p' \over P} ={\rho' \over \rho} + {\theta \over T} + O(\rho' \, \theta)
\;,
\label{AD1}
\end{eqnarray}
that follows from the equation of state for ideal gas. We also take into account that characteristic spatial scales for fluctuations of fluid pressure, temperature and density are much less than those for the mean fields.

In stratified turbulence with turbulent heat flux, the correlation function $\langle \theta({\bm x}) \theta({\bm y}) \rangle$ is much larger than the correlation functions of density-density fluctuations or density-temperature fluctuations.
Indeed, the correlation function $\langle \big[\bec{\nabla}^2 \theta({\bm x})\big] \, \bec{\nabla}^2 \theta({\bm y}) \rangle$ is caused by the turbulent heat flux, i.e., $\langle \theta({\bm x}) \, \theta({\bm y}) \rangle \propto - \tau_0  \, \langle u_i({\bm x}) \, \theta({\bm y}) \rangle \, (\nabla_i T)$, where $\tau_0$ is the characteristic turbulent time. On the other hand, the correlation functions of density-density fluctuations or density-temperature fluctuations are nearly independent of the turbulent heat flux, and they are proportional to the mass flux $\langle {\bm u}({\bm x}) \, \rho'({\bm y}) \rangle$, which
is very small for low Mach number flows. In particular, the temperature fluctuations can be estimated as $\theta \propto - \tau_0 \, u_i \, \nabla_i T$. Consequently, the temperature-density correlator can be estimated as $\langle \theta({\bm x}) \, \rho'({\bm y}) \rangle \propto - \tau_0  \, \langle u_i({\bm x}) \, \rho'({\bm y}) \rangle \, (\nabla_i T)$. The density fluctuations are determined by the continuity equation:
\begin{eqnarray}
{\partial \rho' \over \partial t} = - \bec{\nabla} {\bm \cdot}  (\rho {\bm u}'+\rho' {\bm U}) + O(\rho' {\bm u}') \;,
\label{AD5}
\end{eqnarray}
where ${\bm U}$ is the mean fluid velocity.
The correlation function of density-density fluctuations $\langle \rho'({\bm x}) \, \rho'({\bm y}) \rangle$ is determined by the following equation
\begin{widetext}
\begin{eqnarray}
{\partial \over \partial t} \langle \rho'({\bm x}) \, \rho'({\bm y}) \rangle &=&  - \rho \, \big[\nabla_i^{(y)} \, \langle \rho'({\bm x}) \, u'_i({\bm y}) \rangle
+ \nabla_i^{(x)} \, \langle \rho'({\bm y}) \, u'_i({\bm x}) \rangle \big] -{\nabla_i \, \rho \over \rho} \, \big[\langle \rho'({\bm x}) \, u'_i({\bm y}) \rangle
+ \langle \rho'({\bm y}) \, u'_i({\bm x}) \rangle \big] \;,
\label{AD6}
\end{eqnarray}
\end{widetext}
\noindent
which follows from Eq.~(\ref{AD5}).
Since $\langle \rho'({\bm x}) \, u'_i({\bm y}) \rangle$ is very small (it is of the order of  O(Ma$^2$), where Ma is the Mach number, see \cite{CH02}), and is nearly independent of the turbulent heat flux, the correlation functions of the density-density fluctuations or density-temperature fluctuations are much smaller than the correlation functions of the temperature-temperature fluctuations,
i.e.,
\begin{eqnarray}
{1 \over T^2} \, |\langle \big[\bec{\nabla}^2 \theta({\bm x})\big] \, \bec{\nabla}^2 \theta({\bm y}) \rangle | \gg {1 \over \rho^2} \, |\langle \big[\bec{\nabla}^2 \rho'({\bm x})\big] \, \bec{\nabla}^2 \rho'({\bm y}) \rangle | \;,
\nonumber\\
\label{AD2}\\
{1 \over T} \, |\langle \big[\bec{\nabla}^2 \theta({\bm x})\big] \, \bec{\nabla}^2 \theta({\bm y}) \rangle | \gg {1 \over \rho} \, |\langle \big[\bec{\nabla}^2 \rho'({\bm x})\big] \, \bec{\nabla}^2 \theta({\bm y}) \rangle | \;,
\nonumber\\
\label{AD3}\\
{1 \over T} \, |\langle \big[\bec{\nabla}^2 \theta({\bm x})\big] \, \bec{\nabla}^2 \theta({\bm y}) \rangle | \gg {1 \over \rho} \, |\langle \big[\bec{\nabla}^2 \theta({\bm x})\big] \, \bec{\nabla}^2 \rho'({\bm y}) \rangle | \; .
\nonumber\\
\label{AD4}
\end{eqnarray}
In ${\bm k}$ space the correlation function $\langle \tau \big[\bec{\nabla}^2 \theta({\bm x})\big] \, \big[\bec{\nabla}^2 \theta({\bm y})\big] \rangle$ reads:
\begin{eqnarray}
&&\langle \tau \big[\bec{\nabla}^2 \theta({\bm x})\big] \, \big[\bec{\nabla}^2 \theta({\bm y})\big] \rangle
\nonumber\\
&& \quad = \int \tau(k) \, k^4 \,  \langle \theta({\bm k}) \, \theta(-{\bm k}) \rangle
\, \exp \big(i {\bm k} {\bm \cdot} {\bm R}\big) \, d{\bm k} .
\label{AL6}
\end{eqnarray}

\end{document}